\documentclass[sigconf]{acmart}
\usepackage{algorithmic}
\usepackage{graphicx}
\usepackage{textcomp}
\usepackage{xcolor}
\usepackage{booktabs}

\setcopyright{acmcopyright}
\copyrightyear{2018}
\acmConference[ITiCSE '23]{ACM Conference on Innovation and Technology in Computer Science Education 2023}{July 10--12, 2023}{Turku, FI}

\copyrightyear{2023}
\acmYear{2023}
\setcopyright{acmlicensed}\acmConference[ITiCSE 2023]{Proceedings of the
2023 Conference on Innovation and Technology in Computer Science Education
V. 1}{July 8--12, 2023}{Turku, Finland}
\acmBooktitle{Proceedings of the 2023 Conference on Innovation and
Technology in Computer Science Education V. 1 (ITiCSE 2023), July 8--12,
2023, Turku, Finland}
\acmPrice{15.00}
\acmDOI{10.1145/3587102.3588789}
\acmISBN{979-8-4007-0138-2/23/07}

\begin{document}

\title{The Name of the Title Is Hope}

\author{Larissa Salerno}
\email{lsalernodeca@student.unimelb.edu.au}
\affiliation{%
  \institution{The University of Melbourne}
  \streetaddress{Parkville VIC 3010}
  \city{Melbourne}
  \state{Victoria}
  \country{Australia}
  \postcode{3000}
}

\author{Simone de França Tonhão}
\email{siimone.franca@gmail.com}
\affiliation{%
  \institution{State University of Maringá}
  \streetaddress{Av. Colombo, 5790}
  \city{Maringá}
  \state{Paraná}
  \country{Brazil}
  \postcode{Anonymous postcode}
}

\author{Igor Steinmacher}
\email{igor.steinmacher@nau.edu}
\affiliation{%
  \institution{Northern Arizona University}
  \streetaddress{S San Francisco St, 86011}
  \city{Flagstaff}
  \state{Arizona}
  \country{United States}
  \postcode{Anonymous postcode}
}

\author{Christoph Treude}
\email{christoph.treude@unimelb.edu.au}
\affiliation{%
  \institution{The University of Melbourne}
  \streetaddress{, 86011}
  \city{Melbourne}
  \state{Victoria}
  \country{Australia}
  \postcode{Anonymous postcode}
}
\renewcommand{\shortauthors}{Larissa Salerno, Simone de França Tonhão, Igor Steinmacher, \& Christoph Treude}

\begin{sloppy}

\title[Barriers and Self-Efficacy]{Barriers and Self-Efficacy: A Large-Scale Study on the Impact of OSS Courses on Student Perceptions}

\begin{abstract}

Open source software (OSS) development offers a unique opportunity for students in Software Engineering to experience and participate in large-scale software development, however, the impact of such courses on students' self-efficacy and the challenges faced by students are not well understood. This paper aims to address this gap by analyzing data from multiple instances of OSS development courses at universities in different countries and reporting on how students' self-efficacy changed as a result of taking the course, as well as the barriers and challenges faced by students.

\end{abstract}

\begin{CCSXML}
<ccs2012>
<concept>
<concept_id>10010405.10010489.10010492</concept_id>
<concept_desc>Applied computing~Collaborative learning</concept_desc>
<concept_significance>500</concept_significance>
</concept>
<concept>
<concept_id>10011007.10011074.10011134.10003559</concept_id>
<concept_desc>Software and its engineering~Open source model</concept_desc>
<concept_significance>500</concept_significance>
</concept>
<concept>
<concept_id>10002951.10003227.10003233.10003597</concept_id>
<concept_desc>Information systems~Open source software</concept_desc>
<concept_significance>500</concept_significance>
</concept>
</ccs2012>
\end{CCSXML}

\ccsdesc[500]{Applied computing~Collaborative learning}
\ccsdesc[500]{Software and its engineering~Open source model}
\ccsdesc[500]{Information systems~Open source software}

\keywords{open source software, barriers, self-efficacy, education}

\maketitle

\section{Introduction and Motivation}

As part of their coursework, students in Software Engineering often do not get the opportunity to participate in large-scale software projects with hundreds of developers, thousands of files, and long project history. Yet, many of the challenges inherent in software development only become apparent when software development is conducted at such a large scale. While it is often unrealistic to embed students in an industry project for a semester, open source software (OSS) development offers a unique opportunity for students to experience and participate in large-scale software development. 

Recognizing this opportunity, several universities are now offering dedicated courses that introduce students to OSS development and guide them in making their first contribution to an open source project. But what is the impact of such courses?

Using data from four instances of three courses at three universities in three countries and a total of 359 students, we report on how students' self-efficacy changed as a result of taking a course on OSS development, what barriers the students expected before starting, and what challenges they actually faced in retrospect.

\section{Related Work}

%\subsection{Related Work}
Several recent efforts studied how OSS projects are used in the context of a classroom~\cite{morgan2014lessons, smith2014selecting, buchta2006teaching, coppit2005large, sarma2016training}. Some aimed to understand how projects used in the classroom are chosen. For example, Smith et al.~\cite{smith2014selecting} focused on selecting the most appropriate projects for students' work.  Morgan and Jensen~\cite{morgan2014lessons} detailed the experience of teaching a Software Engineering course based on OSS projects.

Other papers report experiences of using OSS in different courses and contexts. Buchta et al.~\cite{buchta2006teaching} reported their experience in teaching software maintenance and evolution aspects in a Software Engineering course. Holmes et al.~\cite{holmes2014lessons, holmes2018dimensions} reported the lessons of their Undergraduate Capstone OSS Projects (UCOSP) in two instances. They present details of the course, benefits, and potential challenges.

Holmes et al.~\cite{holmes2018dimensions} also analyzed how students perceived the opportunity of taking the capstone course based on OSS. They report that students took advantage of the opportunity to apply their skills in real tasks, from real projects, while receiving real feedback from project maintainers. Steinmacher et al.~\cite{steinmacher2016overcoming} also report the perception of students who contributed to OSS projects as part of an undergraduate course. They were interested in understanding the impact of a portal on the students' perceived self-efficacy. 
Similarly, Pinto et al.~\cite{pinto2019training} investigated the perspective of students a few years after they took a course based on contributions to OSS projects. They found that students recurrently report challenges from social, process, and technical natures but they also report benefits related to improving their technical skills and their self-confidence. 

Differently from the previous literature, in this paper we take a closer look at how an OSS course may change the perception of students about contributing to an OSS project. We aim to understand the shift in terms of self-perceived efficacy and in terms of barriers expected and actually faced during the contribution process.

\section{Course Design}

The courses considered in this study were Software Engineering courses with a focus on software processes; the courses had been taught by two of the authors at three institutions in three different countries, but all followed a similar structure. In class, students learned about OSS development practices, tools, processes, the history of OSS development, licenses, and research on newcomer onboarding and mining software repositories. These were taught through lectures and exercises on topics such as source code management, code review, and continuous integration using GitHub. 

% During class time, students were introduced to the fundamentals of OSS development practices, tools, and processes. This was done through lectures and exercises on topics such as source code management, code review, and continuous integration using GitHub as the learning platform. The lecture material included a history of OSS development and a discussion of different licenses and their consequences. Near the end of the course, lectures covered research related to OSS development, including newcomer onboarding and mining software repositories.

For assessment, students were required to complete online quizzes on the theoretical lecture material and individual ``mini-netnographies'' in which they analyzed the progression of selected GitHub users from their first contribution to their current role in OSS development. This was intended to introduce students to role models in the field. The majority of the assessment was focused on team projects, in which students worked in groups of approximately five to make a contribution to an open source project. The lecturers provided a selection of projects for the student teams to choose from and contacted the project maintainers in advance to ensure that student contributions would be received in a timely and respectful manner. In most cases, the lecturers already had an ongoing relationship with these maintainers from previous collaborations.

Each student team was tasked with selecting a non-trivial open issue from an open source project and developing a plan for addressing it. The lecturers provided feedback on these plans through a short team presentation to guide the teams as needed. The teams then had a few weeks to complete their proposed open source contribution and were encouraged to submit an initial pull request early to allow for multiple rounds of feedback from the project maintainers. The final assessment was based on team presentations and the submitted pull requests. Most of the marks were awarded based on how well each team followed open source contribution processes and interacted with the project maintainers. Whether the pull request was successfully merged played a secondary role in the grading, as it can depend on factors outside of students' control.

\section{Research Methods}

In this section, we detail data collection and analysis.

%\subsection{Research Questions}

%\textbf{RQ1.} What is the impact of OSS courses on students' perceived self-efficacy?

%\textbf{RQ2.} How do expected and actually experienced challenges compare? 

\subsection{Data Collection}

%The research questions posed in this paper were answered based on the analysis of quantitative and qualitative data collected directly from the students. 
Table~\ref{tab:survey} presents the open and close-ended questions used to collect data from the 359 students who took one of the open source courses taught at one of the three institutions considered in this study.

\begin{table*}[]
\centering
\caption{Survey Questions}
\vspace{-3mm}
\footnotesize
\begin{tabular}{p{13.6cm}ll}
\toprule
\textsc{Before}&                                                                                                    \\
What challenges do you expect to encounter when trying to make a source code contribution to an open source project?&Open-Ended \\
\midrule
\textsc{After}&                                                                                                                          \\
What challenges did you encounter when trying to make a source code contribution to an open source project?&Open-Ended          \\
\midrule
\textsc{Before and After}&                                                                                                               \\
I feel comfortable asking for help from the open source community using electronic communication means.&Likert-Scale        \\
I can write my questions and understand answers in English.&Likert scale & Social                                                    \\
I am good at understanding code written by other people.&Likert scale  & Social                                                      \\
I feel comfortable with the process of contributing to an open source project.& Likert scale  & Process                                \\
I think that contributing to an open source software project is an interesting activity.&Likert scale    & Process                     \\

I feel I can set up and run an application if a set of instructions is properly given.& Likert scale   & Process                       \\
I can choose an adequate task to fix if a list of tasks is given.& Likert scale    & Process                                           \\

I am pretty good at searching for solutions and understanding technical issues by myself.& Likert scale & Technical                      \\
I have pretty good skills to write and change code.&Likert scale & Technical                                                        \\

I can find the piece of code that needs to be fixed given a bug report presenting the issue.& Likert scale  & Technical                  \\

\bottomrule
\end{tabular}
\label{tab:survey}
\end{table*}

First, all students from all courses were asked to answer an open-ended question about the challenges that they anticipated facing \textbf{before} they began their team projects. The same question was asked \textbf{after} their contribution attempt, to provide an account of the challenges they actually encountered. 

Using the same \textbf{before-after} design, we administered a self-efficacy questionnaire with a five-point Likert-scale (Strongly Disagree, Disagree, Neutral, Agree, Strongly Agree) to measure the impact of the course. Self-efficacy  is  a  measure  of  the  confidence  in  the participants’ perceived ability to perform a task, which can impact one’s actual ability to complete a task~\cite{bandura1986explanatory}. The questionnaire applied was borrowed from Steinmacher et al.~\cite{steinmacher2016overcoming}. The items had been further classified into Social, Process, and Technical categories~\cite{steinmacher2018let}, to enable a better understanding of these dimensions in the perceived experience.

\subsection{Quantitative Data Analysis}

To analyze the impact of the OSS course on the students' self-efficacy, we first mapped the Likert scale answers to an ordinal (numeric) scale, from 1 to 5 (with 1 representing Strongly Disagree and 5, Strongly Agree). We kept only the entries from those students who provided answers both \textbf{before} and \textbf{after} the course (n=359).

We, then, mapped a mixed-effects logistic regression model. We used the \textit{answer} provided per question as our dependent variable and the item \textit{type} (social, process, or technical) and \textit{when} the answer was provided (\textbf{before} or \textbf{after} the course) as fixed effects. We also modeled the \textit{participant} and the \textit{item} itself as random effects. 

We used ANOVA to evaluate the differences between \textbf{before} and \textbf{after} answers per type. We used the estimated marginal means (EMMs) to compare among groups after fitting the model and reporting the effect size per item type.

%%%\textit{M = Answer \tilde (when+item\_type) + (1\big|participant) + (1\big|item)}

\subsection{Qualitative Data Analysis}

We analyzed the open-ended answers using the Thematic analysis~\cite{braun2012thematic} approach. Thematic analysis is a method of analyzing qualitative data that involves breaking down and coding the text into meaningful themes. It is useful for understanding how different pieces of information are related, identifying patterns in responses, and uncovering underlying meanings within a dataset. Thematic analysis provides researchers with an effective way to explore their data more deeply by allowing them to identify key ideas or concepts that can be used as the basis for further research or decision-making.

The analysis consisted of two phases, one phase to analyze answers to the question related to challenges the students anticipated \textbf{before} contributing, and another phase for the question about the challenges that students faced \textbf{after} contributing to an OSS project. Two of the authors worked individually on thematically analyzing the challenges reported by the students. The researchers analyzed each answer and derived themes according to the content of each student's response. For simplicity's sake, we decided to adopt the same themes and categories nomenclature from two similar studies that one of the authors conducted in the past ~\cite{steinmacher2019overcoming, balali2018newcomers}; some of the categories did not have a correspondent, so for those we added new terms to the original nomenclature. The outcome of the analysis was a list of themes that were placed into five categories: Newcomers' orientation, Newcomers' characteristics, Communication, Documentation problems, and Technical hurdles.

In total, 265 students responded the \textbf{before} questionnaire, and 191 students answered the \textbf{after} questionnaire. We analyzed the data combining the three courses. The themes were generated according to the content of the students' answers. The researchers read each response carefully, derived a list of challenges reported in each answer, and then generated themes by merging challenges. For example, one participant mentioned $-$ \textit{``Learning the syntax and language that the project uses might take awhile depending on the language and application towards that project''}. From that chunk of text, we identified that the need to learn a new programming language might be a challenge when contributing to an OSS project for the first time, so we added ``Learn a new programming language'' to the list of potential themes. In general, the two researchers had no difficulty in reaching a consensus, as the themes each researcher identified were similar.

%\section{Findings}

\section{Impact on self-efficacy} 

The distribution of answers \textbf{before} and \textbf{after} look very similar when we observe the boxplots in Figure~\ref{fig:boxplot}. However, it is possible to observe a small shift in the mean of the answers after the course. Since visual inspection did not highlight any clear pattern, and since the scale used was small, we focus on the results of the logistic model to understand if there were any trends.

First, as shown in Table~\ref{tab:randomEffs}, the result of the regression showed that the fixed-effects explain more than 50\% of the answers provided by the participants. Residual is greater than the variance explained by the random effects, although the individual preferences (participants) are non-negligible. Analyzing the result of the ANOVA test, we observed that the F-value indicates significant differences for both the time (before and after) and the type of the items (Social, Process, and Technical) -- F-values$=$47.723 and 5.785, respectively.

\begin{figure}
    \centering
    \includegraphics[width=0.48\textwidth, trim={6cm 5.5cm 6cm 5.5cm},clip]{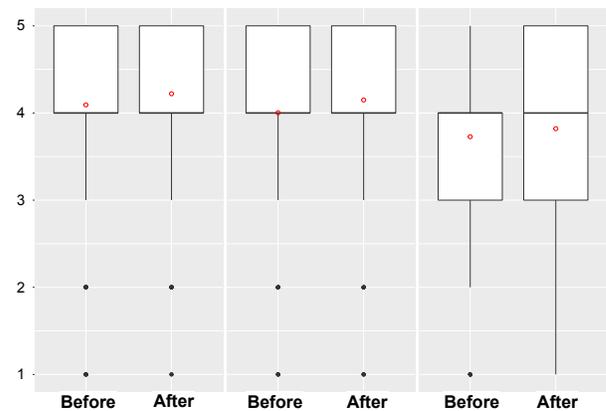}
    \caption{Distribution of answers before and after the course, with answers related to Process, Social, and Technical shown from left to right. The red circle identifies the values average.}
    \label{fig:boxplot}
    \vspace{-5mm}
\end{figure}

\begin{table}[!ht]
    \caption{Random Effects analysis}
    \label{tab:randomEffs}
    \centering
    \footnotesize
    \begin{tabular}{lll}
    \toprule
        Group Name & Variance & Std.Dev. \\ \midrule
        participant (Intercept) & 0.25536 & 0.5053 \\
        item (Intercept) & 0.02632 & 0.1622 \\ 
        Residual & 0.51727 & 0.7192 \\ \bottomrule
    \end{tabular}
\vspace{-1mm}
\end{table}

%\begin{table}[!ht]
%    \caption{Fixed Effects}
%    \centering
%    \begin{tabular}{llll}
%    \toprule
%        ~ & Estimate & Std. Error & t value \\ \midrule
%        (Intercept) & 4.21567 & 0.08686 & 48.536 \\ 
%        whenBefore & -0.11727 & 0.01698 & -6.908 \\ 
%        typeSocial & -0.08113 & 0.14241 & -0.570 \\ 
%        typeTechnical & -0.38301 & 0.11628 & -3.294 \\ \bottomrule
%    \end{tabular}
%\end{table}

\begin{table}[!ht]
    \caption{Results for the ANOVA analysis}
    \centering
    \footnotesize
    \begin{tabular}{llll}
    \toprule
        ~  & Sum Sq & Mean Sq & F-value \\ \midrule
        when  & 24.6854 & 24.6854 & 47.723 \\ 
        type & 5.9848 & 2.9924 & 5.785 \\ \bottomrule
    \end{tabular}
    \vspace{-3mm}
\end{table}

When digging deeper into the differences before and after per dimension, the result of the regression showed a significant difference when comparing the answers provided before and after the course for all three dimensions (Social, Process, and Technical) – p-value<0.001. The effect size showed an increase in the values of answers received after the course for all dimensions ($\approx$0.16, small effect size). Given these results, we conclude that, although with a small effect size, the students perceived themselves as more confident with the OSS contribution process at the end of the courses (the small shift in the mean ---Figure 1--- is indicative of this improvement).

% When investigating differences \textbf{before} and \textbf{after} per dimension, the result of the regression showed a significant difference when comparing the answers provided \textbf{before} and \textbf{after} the course for all three dimensions (Social, Process, and Technical) -- p-value$<$0.001. The effect size showed an increase in the values of answers received \textbf{after} the course for all dimensions ($\approx$0.16, small effect size). Given these results, we conclude that, although with a small effect size, the students perceived themselves as more confident with the OSS contribution process at the end of the courses (the small shift in the mean ---Figure~\ref{fig:boxplot}--- is indicative of this improvement).

\section{Before-After Comparison}

\begin{figure*}
\centering
  \includegraphics[width=1\textwidth]{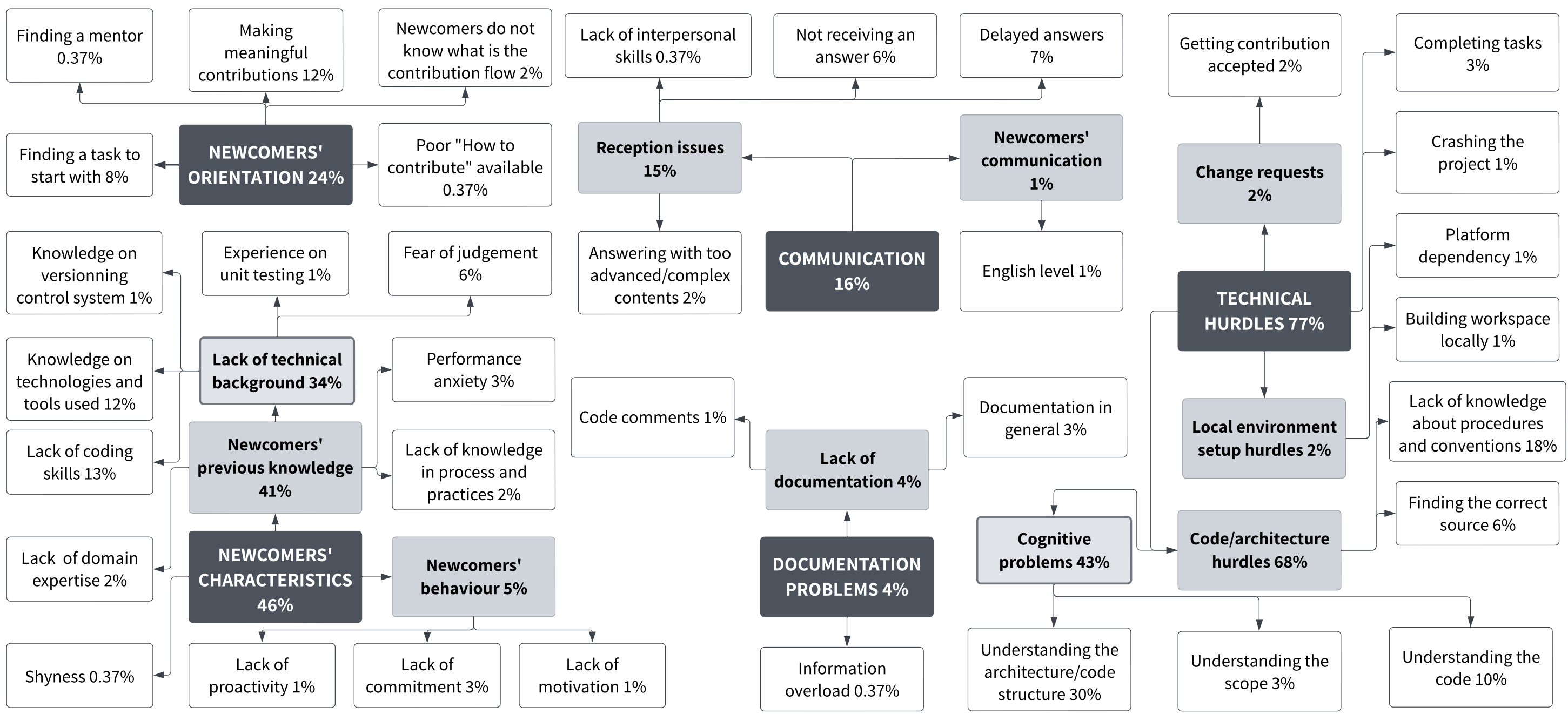}
    \caption{Challenges the students expected before contributing to an open source project.}
   \label{fig:before}
\end{figure*}

We received a total of 265 responses for the \textbf{before} contributing question and  191 answers for the \textbf{after} contributing question. The diagrams in Figure \ref{fig:before} and Figure \ref{fig:after} represent the challenges we identified through the analysis process. We used the categories names represented in similar studies \cite{steinmacher2016overcoming, steinmacher2019overcoming} to group the challenges. The diagram comprises five categories with multiple subcategories, representing the challenges students have reported. The approximate percentage of students who reported each challenge is displayed next to the categories and subcategories' names. 

Due to the large number of subcategories that each category holds, we will focus on discussing the challenges with a higher percentage. The categories will be discussed in separate subsections, presenting the \textbf{before} and \textbf{after} results.

\subsection{Newcomers' Orientation}

Analyzing the data, we identified a number of challenges students expected to find \textbf{before} contributing to an open source project. In total, 24\% mentioned that the orientation and support they received from the community was a key factor. 

Out of the five subcategories, "Making a meaningful contribution" and "Finding a task to start with" were the most prominent, with 12\% of students indicating that not being able to make a meaningful contribution to the project could be a challenge. One of the participants said $-$ \textit{"I think the big challenge is when I have to create a new useful feature for the project. I need to define goals and objectives"}. Finding a task to start with is also an aspect that students mentioned, 8\% stated that the process of picking an issue to work on could be challenging, especially because of their skill level. 

%\begin{figure*}[ht!]
 %   \centering
  %  \includegraphics[width=1\textwidth]{BEFORe3.jpg}
   % \caption{Challenges the students expected before contributing to an Open source project.}
   % \label{fig:before}
%\end{figure*}

\textbf{After} contributing to an open source project, 29\% of students encountered challenges related to the orientation they received. The percentage slightly increased, but the most noticeable change is in the subcategories. Only 2\% of the students stated that making meaningful contributions was challenging, but 16\% mentioned that finding a task to start with was difficult. The main reason why is that they could not predict the task difficulty level. 
One of the participants mentioned mentioned how their team struggled to pick the right issue to work on $-$ \textit{"My team had trouble choosing an issue because we have no idea how difficult one issue is"}, another student mentioned $-$ \textit{"Identifying the "good first issue" for our team to take up. There were numerous issues on the [project] GitHub repository. We wanted to make sure we do not pickup [sic] too hard or too easy to solve issue as the contribution should be significant enough"}.

%\textit{"Identifying the "good first issue" for our team to take up. There were numerous issues on the [project] GitHub repository. We wanted to make sure we do not pickup [sic] too hard or too easy to solve issue as the contribution should be significant enough"}.

\subsection{Newcomers' Characteristics}

\textbf{Before} making a contribution to an OSS project, 46\% of the students expected that their characteristic traits could potentially be a challenge. In terms of their previous knowledge, 34\% of the students believe that their lack of technical background could be an issue, especially when it comes to their lack of coding skills and knowledge of technologies and tools used in the project.

The lack of technical background proved to be a significant hurdle after their contribution. Approximately 22\% of the students faced challenges regarding the technologies and tools used in the projects. The need to learn new programming languages and use tools they had never used before was the main aspect reported by students $-$ \textit{"Before I worked on this project, I knew nothing about JavaFX, and I have not used Gradle once. It was hard for me to learn a new thing from scratch, especially JavaFX since it is a minority framework that not too many people are using it"} said the student. On the other hand, the fear of students regarding their lack of coding skills did not become a reality, as only 3\% of the students reported facing challenges related to this aspect.

% their lack of coding skills did not appear to be a significant problem, as only 3\% of the students reported facing those challenges.

%Before I worked on this project, I knew nothing about JavaFX, and I have not used Gradle once. It was hard for me to learn a new thing from scratch, especially JavaFX since it is a minority framework that not too many people are using it

\subsection{Communication}

\begin{figure*}
    \centering
    \includegraphics[width=0.95\textwidth]{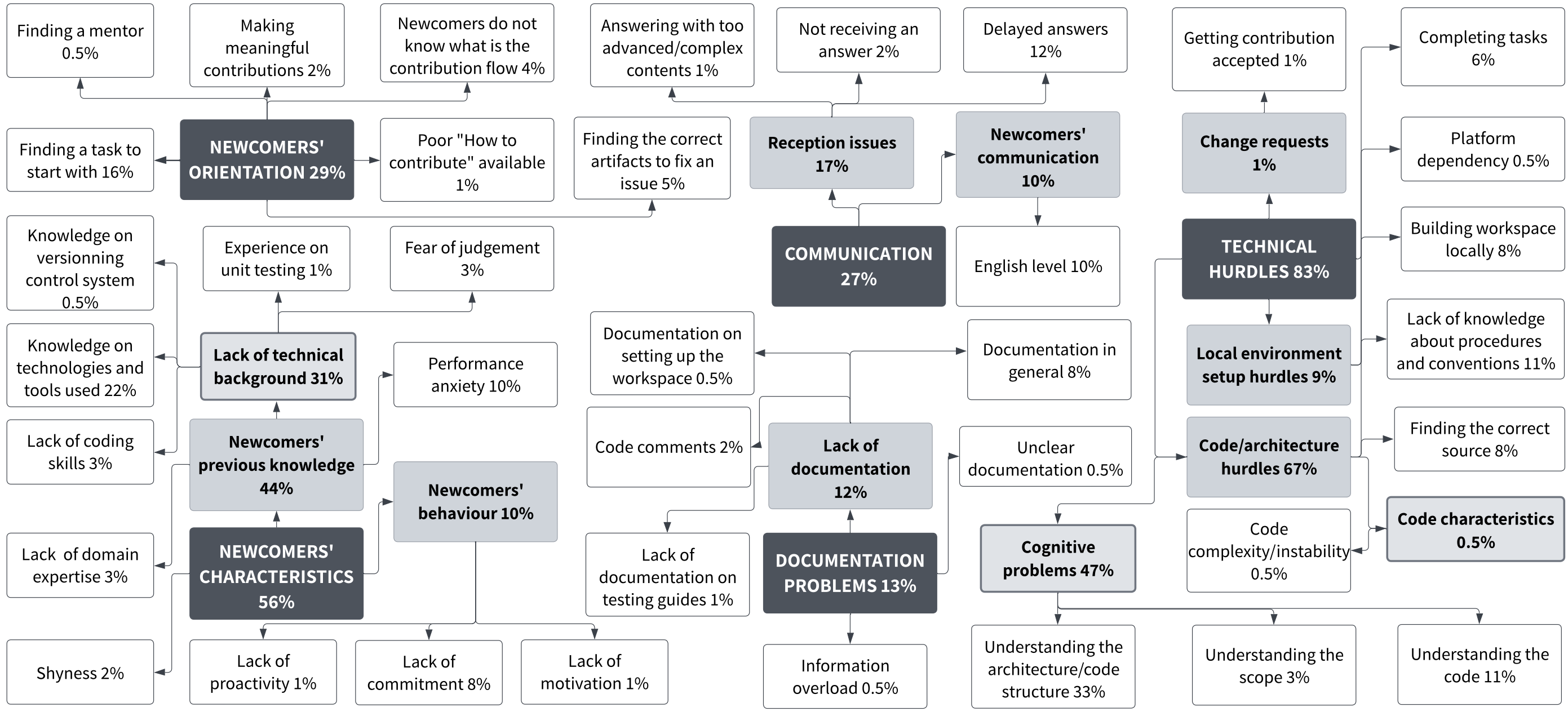}
    \caption{Challenges the students encountered after contributing to an open source project.}
    \label{fig:after}
    \vspace{-3mm}
\end{figure*}

Regarding communication barriers, only 16\% of the participants expected to face any problems in this sense; most participants were expecting to face reception issues, which included receiving delayed answers and not receiving an answer from the community. After contributing, 27\% of the participants encountered communication challenges; 12\% of participants mentioned that they received delayed answers and 2\% did not receive any answer or feedback $-$ \textit{"Our final challenge was getting a response from the Project owner. Although we made a PR, we never got the feedback"}.

The participants' English level was also reported as an issue, as the majority of students are from non-English speaking countries $-$ \textit{"Since most of the open Source projects are English, the language is an important question. Sometimes I can't correctly get the idea about what should I do. I'm confused about my goal. It troubles me deeply".}

\subsection{Documentation Problems}

As the students were having contact with the projects for the first time, the documentation to understand the project and the code was crucial. \textbf{Before} contributing, only 4\% of the participants were expecting to find a lack of documentation about the projects they would be working on. However, this scenario changed \textbf{after} contributing, when 12\% of the participants mentioned that one of the challenges they faced was the lack of documentation about the project. The lack of documentation in general, comments in the code, testing guides, and setting up the environment documentation were aspects they indicated as challenging \textbf{after} contributing.

\subsection{Technical Hurdles}

The category of technical hurdles emerged as the most frequently cited by the students both \textbf{before} and \textbf{after} contributing. \textbf{Before} contributing, 77\% of the students were expecting to face technical challenges, such as challenges related to the understanding of the code structure and architecture of the project. 

Approximately 30\% of students expected challenges when understanding the architecture of the project and code structure $-$ \textit{"Some source code can be very hard to read depending on the structure. The code might be separated into different files with dependencies from another file. Understand what the code is already doing can take some time before making any contribution to the code"}; another student said $-$ \textit{"Understanding the code structure could take quite a while due to either an unusual code style or the size of the project"}.

%; another student also said $-$ \textit{"Understanding the code structure could take quite a while due to either an unusual code style or the size of the project"}. 

Besides the architecture and code structure, 10\% of the students were expecting to face challenges in understanding the code itself, especially in terms of understanding different coding styles $-$ \textit{"The main challenge I would encounter is the steep learning curve that comes with understanding code written by other people. I find the starting point to be the most difficult in every project"}. One student also pointed out how the variety of coding styles can affect the code readability $-$ \textit{"As the community grows and a lot of developers participate, it will be challenging to establish coding standards, as each individual has their own style in coding. This might impact the readability of the code"}. 

The lack of knowledge about procedures and conventions was also a concern for 18\% of the students \textbf{before} contributing. Students believe that not meeting the projects' code standards could prevent them from having their contribution accepted $-$ \textit{"It's possible to receive some negative feedback from the community if my coding practice does not meet the standard"}; another student also shared the same perception $-$ \textit{"Another challenge will be adhering to the requirements of the submission such as some projects might have a certain code coverage that needs to be added or to ensure that all existing test cases pass"}.

The concerns students had regarding technical hurdles became a reality, as 83\% of the students faced technical challenges while making contributions to the projects. Understanding the code structure and architecture was the major challenge, being mentioned by 33\% of the students $-$ \textit{"As expected, understanding the project structure is hard. I was at a loss for how to start at the beginning"}; another student reported $-$ \textit{"Jumping on board and knowing nothing about the source code or the software architecture was quite challenging"}. 

Understanding the code was also a challenge for 11\% of the students, similarly to \textbf{before} contributing. The main issue was related to the different coding styles $-$ \textit{"The coding style was inconsistent across the files of the project. As a result, our team had to take more time trying to figure out which coding style was the most common"}. The lack of knowledge about procedures and conventions also happened to be an issue for 11\% of the students. According to one of them, the project demanded a strict standard, but it was not offering any information or instructions $-$ \textit{"Abiding by coding standards/style of the open source project. We had a few issues where our pull request was not accepted due to the way we did the task"}.

\section{Discussion}

In this section, we discuss the implications of our results for educators as well as threats to the validity of our study.

\subsection{Implications}

\textbf{A single contribution experience is not sufficient.} Our study revealed that a single OSS course might not be enough to improve students' self-efficacy and ability to overcome barriers. Our findings showed that some of the challenges students anticipated turned out to be true, while others were even more difficult to overcome than expected. For instance, we observed an increase in performance anxiety, from 3\% expected to 10\% encountered. To address this, we recommend educators consider incorporating OSS contributions into multiple courses and/or encouraging participation in programs such as Google Summer of Code, despite scheduling constraints.

\textbf{Tools and technologies are crucial for success.} Our study found that students were initially worried about their lack of coding skills (13\%) and knowledge of tools and technologies (12\%) \textbf{before } starting their OSS contribution journey. However, in hindsight, they realized that knowledge of tools and technologies (22\%) was a much greater issue than coding skills (3\%). While coding skills are necessary, we recommend educators to incorporate tools and technologies into their curriculum. Real-world software projects, both in industry and open source, rely heavily on tools and technologies, and students' proficiency in using them is crucial for their future.

\textbf{Documentation issues are more prevalent than anticipated.} One of the most striking differences in our ``\textbf{before}'' and ``\textbf{after}'' survey responses was related to documentation problems. While only 4\% of the students expected such issues, they were actually encountered by 13\%. The problem of inadequate or outdated documentation is a well-established issue in Software Engineering literature. Therefore, we urge educators to prepare students for the realities of software documentation, including teaching them how to write clear and comprehensive documentation and how to navigate code bases where documentation may be lacking.

\textbf{Non-native speakers face difficulties with conventions, communication, and documentation.} Large-scale software development is equally about communication and collaboration as it is about programming. English is the primary language of communication in most projects, and students for whom it is not their first language can struggle with understanding and adapting to common conventions and styles. Educators can support these students by providing them with templates for effective first messages to a project or pull request titles and descriptions. Encouraging them to study and learn from successful contributions by other open source contributors can serve as a guide for them to communicate effectively with contributors from diverse backgrounds.

%\textbf{Frustration when tasks are already solved by other contributors.} A unique challenge in making open-source contributions is that other contributors in the ecosystem may provide a bug fix or feature enhancement before students have a chance to do so. In industry settings, this inefficiency can be avoided through task management, like task boards in Agile settings. However, this issue is not common in university settings since many assignments require students to submit solutions to the same problem. To mitigate this challenge in open-source settings, educators should coordinate with OSS projects to reserve specific tasks for students, and carefully organize teams to prevent overlap. However, it is important to note that matching newcomers to tasks that are appropriate for their skill level is not a straightforward task.

\textbf{Understanding code and code structure requires significant effort.} This includes becoming familiar with coding styles, conventions, and best practices used in the project. However, the strict time frame of a university course does not align well with the flexible nature of OSS contributions. OSS projects are ongoing and can take a long time to fully understand, whereas university courses are often limited to a specific semester or term, creating a mismatch between the two environments. Educators should be aware of this mismatch and provide students with adequate time and resources to become proficient in navigating and contributing to OSS projects \textbf{before} they are ready to work on their first contribution.

\subsection{Threats to Validity}

The conclusions we make are based on data from 359 students from four instances of three courses at three universities in three countries. While we consider this to be a large sample size, we note that our findings may not necessarily generalize to other courses or student populations. Additionally, the scope of the projects that the students contributed to is relatively limited, and primarily comprises projects that the lecturers were familiar with. This may not accurately reflect the experience of students who work on projects that are not specifically advised to expect student contributions. Furthermore, the qualitative analysis component of our research introduces an element of subjectivity. To address this concern, we employed a nomenclature in line with similar studies in the field.

The survey responses may be influenced by social desirability bias, where participants provide responses that they believe are socially acceptable rather than accurate. To mitigate this threat, we emphasized to students that their answers would not affect their grades. Additionally, the measures used to assess self-efficacy may not be completely reliable or valid. We followed established practices in phrasing the self-efficacy questions in the survey.

\section{Conclusion}

This paper aimed to address the gap in understanding the impact of OSS development courses on students' self-efficacy and the challenges faced by them. Through analyzing data from multiple instances of OSS development courses at universities in different countries, we found that students' self-efficacy slightly improved as a result of taking the course. Additionally, we identified that many of the challenges anticipated by students actually occurred, with issues related to tools, technologies, and documentation being more prevalent than expected. Based on these findings, we provide implications for educators on how to best guide students to make successful contributions to an OSS project.

Future research in this area could aim to better understand the long-term effects of participating in open-source software development on students' careers and professional growth. Additionally, it would be interesting to explore how the students' Software Engineering skills were impacted by their experience with the OSS course as well as investigate the differences between the courses in more detail. Thus, it would be valuable to explore and develop effective strategies and best practices for both OSS projects and educators to support and guide students through any challenges and barriers they may encounter during their participation in OSS development. This could not only enhance students' learning experiences, but also increase the chances of successful contributions to open-source projects, ultimately promoting both students' education and the sustainability of OSS projects.

\section*{Acknowledgment}

This work is partially supported by the National Science Foundation under Grant number 2247929.

\bibliographystyle{ACM-Reference-Format}
\bibliography{seet23-selfefficacy}

\end{sloppy}
\end{document}